\newif\ifreview
\newif\ifgrayscale

\reviewfalse

\grayscalefalse

\ifreview
\documentclass[review,number,sort&compress,letter]{elsarticle}
\usepackage{lineno}
\linenumbers
\else
\documentclass[twocolumn,final,sort&compress,letter]{elsarticle}
\fi

\usepackage{units}
\usepackage{amssymb,amsmath}
\usepackage[titletoc]{appendix}
\usepackage{hyperref}
\usepackage{graphicx}
\usepackage{tabu}
\ifgrayscale
\PassOptionsToPackage{gray}{xcolor}
\usepackage{tikz}
\fi
\usepackage[siunitx]{circuitikz}
\usepackage{mhchem}

\newcommand{\exclude}[1]{}
\newcommand{\superbkerned}{Super\kern-.1em\textit{B}\kern.2em}
\newcommand{\superb}{Super\kern-.1em\textit{B}}
\ifgrayscale
\newcommand{\smartpath}[1]{figures_grayscale/#1}
\else
\newcommand{\smartpath}[1]{figures_colour/#1}
\fi

\begin{document}
\begin{frontmatter}

\title{Improved Particle Identification Using Cluster Counting in a Full-Length Drift Chamber Prototype}

\date{\today}

\author[ubc]{Jean-Fran\c{c}ois Caron\corref{cor1}}
\ead{jfcaron@phas.ubc.ca}
\author[ubc]{Christopher Hearty}
\author[ubc]{Philip Lu}
\author[ubc]{Rocky So}

\author[mcgill]{Racha Cheaib}

\author[montreal]{Jean-Pierre Martin}

\author[TRIUMF]{Wayne Faszer}

\author[uvic]{Alexandre Beaulieu}
\author[uvic]{Samuel de Jong}
\author[uvic]{Michael Roney}

\author[LNF]{Riccardo de Sangro}
\author[LNF]{Giulietto Felici}
\author[LNF]{Giuseppe Finocchiaro}
\author[LNF]{Marcello Piccolo}

\address[ubc]{The University of British Columbia,
6224 Agricultural Road,
Vancouver, BC, 
Canada V6T 1Z1}
\address[mcgill]{McGill University,
3600 rue University,
Montreal, QC,
Canada H3A 2T8}
\address[montreal]{Universit\'{e} de Montr\'{e}al,
2900 Boulevard \'{E}douard-Montpetit,
Montr\'{e}al, QC,
Canada H3T 1J4}
\address[TRIUMF]{TRIUMF,
4004 Wesbrook Mall,
Vancouver, BC,
Canada V6T 2A3}
\address[uvic]{University of Victoria,
PO Box 3055, STN, CSC,
Victoria, BC,
Canada V8W 3P6}
\address[LNF]{Laboratori Nazionali di Frascati dell'INFN,
Via Enrico Fermi 40,
I-00044 Frascati,
Italy}

\cortext[cor1]{Corresponding author.}
\cortext[cor1]{\textit{Telephone number:} +1 604 822-1445}

\begin{abstract}
Single-cell prototype drift chambers were built at TRIUMF and tested
with a $\sim\unit[210]{MeV/c}$ beam of positrons, muons, and pions.  A
cluster-counting technique is implemented which improves the ability
to distinguish muons and pions when combined
with a traditional truncated-mean charge measurement.  Several
cluster-counting algorithms and equipment variations are tested, all
showing significant improvement when combined with the traditional method.  The
results show that cluster counting is a feasible option for any particle
physics experiment using drift chambers for particle identification.
The technique does not require electronics with an overly high
sampling rate.  Optimal results are found with a signal smoothing time
of $\sim\unit[5]{ns}$ corresponding to a $\sim\unit[100]{MHz}$ Nyquist frequency.
\end{abstract}


\end{frontmatter}
\tableofcontents
\section{Introduction}
This paper describes the development and testing of a prototype drift
chamber whose purpose is to evaluate the feasibility of a
``cluster-counting''
technique\cite{Walenta_TEC} for implementation in a high luminosity
$e^+e^-$ experiment.  
Cluster counting is expected to
improve particle identification (PID) by reducing the effect of
fluctuations in drift chamber signals.  These are due to gas amplification and the
fluctuation in the number of primary electrons per ionization site.  There
may also be improvements in tracking resolution, but this is left for
a later study. The
requirement of fast electronics and larger data sizes may make the
technique impractical in terms of capital costs, available space
near the detector, and computing power.  To date the technique has not been deployed
in an operating experiment.  
This work demonstrates that a cluster-counting drift chamber is a
feasible option for an experiment such as \superb
\cite{SuperB_CDR}\cite{2010arXiv1007.4241S}.  \superbkerned
was cancelled after the experiments described in this paper, but the results are applicable to any drift
chamber that is used for particle identification.  The design
of our prototype chambers was strongly influenced by
the demands of \superbkerned, which are described in the Technical
Design Report\cite{2013arXiv1306.5655S}.

\subsection{Drift Chambers}
Drift chambers are general-purpose detectors that can track and
identify charged particles\cite{Charpak1968262}\cite{blum2008particle}.  They consist of a large volume of gas
with instrumented wires held at different voltages.  When charged
particles move through the chamber they ionize the gas particles.  The electrons
from these primary ionizations drift towards the wires held at high
positive voltage, while the ions drift towards the grounded wires.
The sense wires are very thin ($\sim \unit[20]{\mu m}$), such that the strong
electric field accelerates the electrons enough to cause further
ionization near the sense wire.  The new electrons ionize further
into an avalanche, which is registered as an electronic signal on the
sense wire.  The amplification of the low-integer number of primary
ionization electrons into a detectable signal on the wire is called the gas
gain.  

The energy loss of  a heavy ($m \gtrsim \unit[1]{MeV/c^2}$) charged particle from
primary ionizations depends on its speed, as given by the Bethe formula\cite{bethe} and various
corrections\cite{PhysRevD.86.010001}.  The speed measurement is combined with the independent
momentum measurement from tracking, giving the particle's mass, which
is a unique identifier.  To measure speed, we measure or estimate a
quantity proportional to the number of primary
ionizations.  A traditional drift chamber accomplishes this by
measuring the total ionization per unit length of the track, which is proportional to the integral of the
electronic signal on the sense wires belonging to a track.  
The theoretical probability distribution function for the total ionization is a Landau distribution, which has an infinite mean and
standard deviation\cite{blum2008particle}.  The consequence is that if one takes the average
of a number of samples (e.g.\ 40 measurements of deposited charge in a
track), the resulting distribution is non-Gaussian and is dependent on
the number of samples taken.  Instead of the mean of the distribution,
one can use the most probable value for the total ionization.  This is
accessed by a truncated mean technique.  Our truncated mean
procedure is described in Sec.~\ref{bootstrapping}.

\subsection{Cluster Counting}
The conventional technique described above is sensitive to gas gain
fluctuations as well as the statistical fluctuations in the number of
primary electrons produced in each ionization event.  Moreover, the
truncated mean procedure that is typically used discards a substantial
fraction of the available information.  None of these disadvantages
exist if the number of primary ionizations can be measured more directly.
\subsubsection{Technique}
The cluster-counting technique involves resolving the cluster of avalanching
electrons from each primary ionization event.  This is done by digitizing the signal from the
sense wire in each cell and applying a suitable algorithm.  The rise time of the
signal from a cluster is approximately $\unit[2]{ns}$, so electronics
with sufficiently high bandwidth are required.  

In principle, clusters can be detected as long
as they do not overlap completely in time.  This can happen
irrespective of the electronics involved due to the probabilistic
nature of the ionization process.  Overlapping clusters are more
likely for highly oblique tracks.  Complex algorithms which consider
signal pulse heights might disentangle even overlapping clusters, but
the algorithms tested in this work do not.

An optimal algorithm would have a high efficiency for identifying true
clusters and a low rate of reporting false clusters (due to noise for
example).  
\subsubsection{PID}\label{CC_PID}
In traditional drift chambers using the
integrated signal, the signal amplitude is determined by the
convolution of the probability of primary ionization, the number of
primary electrons produced, and the variations in gas gain.  This
results in a long-tailed distribution that is typically dealt with by the truncated
mean procedure.  Conversely, if clusters are perfectly identified, then the only variation
is from the primary ionization, which is a Poisson process.  No
cluster counts need to be discarded to allow for a proper statistical
treatment.  In reality some counted clusters will be missing or fake,
the rate of these being caused by gas gain fluctuations, noise level,
and the time separation capabilities of the electronics.  The idea is
that the sensitivity to these effects is small.  The difficulty arises from the need to optimize an
arbitrarily complex cluster-counting algorithm. 

A difficulty with both charge integration and cluster counting is
the presence of $\delta$-rays\cite{blum2008particle}.  These are electrons produced in
primary ionizations that travel far in the gas before
further ionizing, such that they create their own separate ionization
cluster.  The production of $\delta$-rays at a given momentum depends only
on the particle speed ($\propto 1/\beta^2$)\cite{PhysRevD.86.010001}.  This
inflates the charge integral and the cluster count with only a weak dependence
on the species of the original particle, the result is a decreased PID
resolution in general.  The presence of $\delta$-rays is one of the
reasons why a truncated mean is used in the charge integration
method.  While cluster counting is also affected by $\delta$-rays, the
effect is less pronounced, allowing all of the data to be used.

\subsubsection{Cluster Timing}
Any cluster-counting algorithm that uses a digitized signal is able to
report not only the number of clusters in a cell, but also the arrival
time of each of those clusters.  
In the oversimplified case of a
linear and homogenous drift velocity and infinite cells, the average
spacing in time between consecutive clusters would simply be
proportional to the inverse of the number of clusters in the cell.
In a more realistic scenario, the average spacing between clusters is
useful information that is not one-to-one with the number of
clusters.  We can exploit the lack of perfect correlation and use the
cluster timing information to further improve our ability to identify particles.
\subsubsection{Tracking}
For tracking, cluster counting may also improve performance, but in a
much lesser degree and more subtle manner than as for PID.  A
traditional drift chamber uses only the arrival time of the overall
signal in determining the distance of closest approach from a sense
wire.  Unfortunately this arrival time measurement is vulnerable to
noise, gas gain fluctuations (small initial clusters may be missed),
etc.  If the first few clusters are resolved, then while the
first cluster arrival time is still the primary datum, the second cluster
arrival time can be used as a consistency check.  If the second
cluster arrives much too late, then the chance that the first cluster
was a fake is greater, so a smaller statistical weight can be assigned
to that cell when reconstructing the whole track.  This paper
deals only with the PID improvements and does not address tracking.

\section{Apparatus}
In this section we describe the prototype drift chambers that were
built, the custom signal amplifiers and the various types of cables
that were tested.  We also describe the experimental setup in the test
beam, the data acquisition system, and the devices used for external
PID and triggering. 

\subsection{Prototype Drift Chambers}
\begin{figure}
  \centering
  \begin{tikzpicture}
    \node[anchor=south west,inner sep=0] (image) at
    (0,0){\includegraphics[width=\columnwidth]{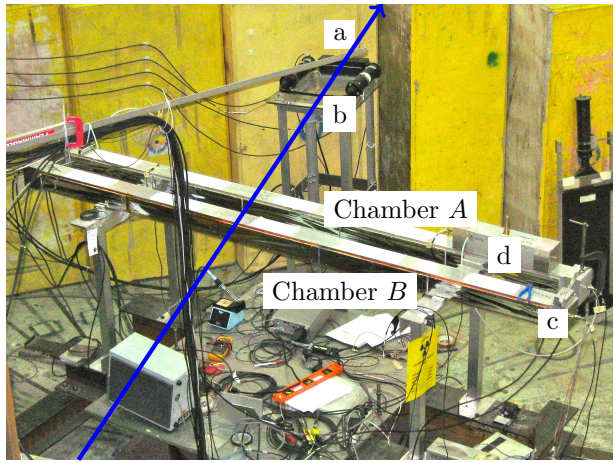}};
    \begin{scope}[x={(image.south east)},y={(image.north west)}]
      \node [fill=white] at (0.65,0.55) {Chamber $A$};
      \node [fill=white] at (0.55,0.37) {Chamber $B$};
      \node [fill=white,above] at (0.55,0.9) {a};
      \node [fill=white,below] at (0.55,0.8) {b};
      \node [fill=white] at (0.9,0.3) {c};
      \node [fill=white] at (0.82,0.45) {d};
      \draw [->,blue, ultra thick] (0.12,0) -- (0.62,1.0);
    \end{scope}
  \end{tikzpicture}
  \caption{Photo of the prototype chambers mounted during our beam
    test.  The far scintillator (Sec.~\ref{TOFsection}) and
    additional PMTs (Sec.~\ref{trigger}) are visible in the
    background (labelled a and b respectively).  The amplifier shielding
    boxes (c) are on the right side of
    the picture.  The smaller monitoring chamber (labelled d) (Sec.~\ref{DAQ}) is
    on top of Chamber B.  The blue arrow shows the path of the particle
    beam through our prototypes and two of the scintillators.}
  \label{M11_Photo}    
\end{figure}

We built two nearly identical full-length ($\unit[2.7]{m}$) single-cell drift
chambers, called chamber $A$ and chamber $B$ (Fig.~\ref{M11_Photo}).
The only difference between the two chambers is the diameter of the
sense wires: $\unit[20]{\mu m}$ for Chamber $A$ and $25$ or
$\unit[30]{\mu m}$ for Chamber $B$.  More details about the wires are
given below.

\begin{figure}
\centering
\includegraphics[width=\columnwidth,clip=true,trim = 0 0 30 52]{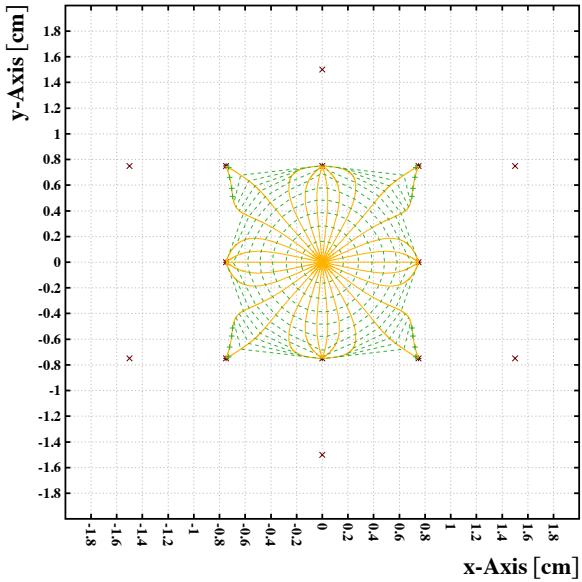}
\caption{Garfield\cite{garfield} simulation of isochrones for electron drift
times in our prototypes, with $90:10$ helium and isobutane.  The
isochrone intervals (dashed lines) are $\unit[50]{ns}$.  The full
orange lines are the drift paths.  The central point is the sense wire
at high voltage, while the 8 points in a square around it are the
field wires at ground.  The extra 6 points outside the cell are bias
wires to simulate the presence of an infinite network of cells.  The
wire voltages are $\unit[1820]{V}$ and $\unit[1054]{V}$ for the sense
wire and bias wires respectively.}
\label{isochrones}
\end{figure}

The wire layout creates a square cell \unit[15]{mm} wide
in a $10\times\unit[10]{cm}$ cross-section casing (for a gas volume of
$\unit[2.7\times 10^4]{cm^3}$).  Figure \ref{isochrones} shows a cell diagram including the dimensions
and wire locations.  The
aluminium casing of the chambers has five large windows on two sides of the cell to
allow particles to enter and exit unimpeded.  The windows are made
of thin ($\sim\unit[20]{\mu m}$) aluminium, protected by aluminized Mylar.

Different amplifiers are mounted on the endplates of the
drift chambers, connected directly to the sense wires.  The amplifiers vary in their gain, input impedance, and
bandwidth.  They are described in more detail in Sec.~\ref{amplifiers}.

\begin{figure}
\centering
\begin{circuitikz}
  \draw (-0.5,0) node[]{HV} (-0.25,0) -- (0,0);
  \draw (0,0) to[R=10<\kilo\ohm>] (2,0);
  \draw (2,0) to[R=1.5<\mega\ohm>] (4,0);
  \draw (4,0) -- (4.25,0);
  \draw (5.25,0) node[]{Sense Wire};
  \draw (2,0) to[C=1000<\pico\farad>] (2,-2);
  \draw (2,-2) node[ground] {};

  \draw (2,0) -- (2,1) to[R=390<\ohm>] (4,1) -- (4,0);
\end{circuitikz}
\begin{circuitikz}
  \draw (-0.5,0) node[]{HV} (-0.25,0) -- (0,0);
  \draw (0,0) to[R=10<\kilo\ohm>] (2,0);
  \draw (2,0) to[R=1.5<\mega\ohm>] (4,0);
  \draw (4,0) -- (4.25,0);
  \draw (5.25,0) node[]{Sense Wire};
  \draw (2,0) to[C=1000<\pico\farad>] (2,-2);
  \draw (2,-2) node[ground] {};
\end{circuitikz}
\caption{Circuit diagram representing the high voltage connection to
  the sense wire with (top) and without (bottom) termination resistor.  Without the $\unit[390]{\Omega}$
  resistor, the signal can bounce.\label{Termination}}
\end{figure}
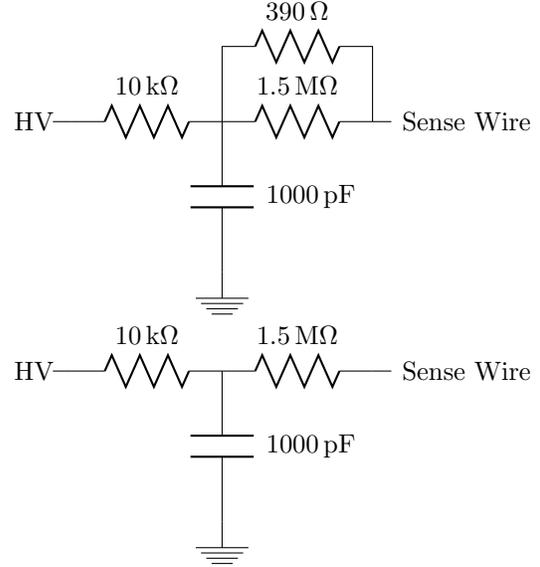

We had the option of including a termination resistor to ground on
the non-instrumented side of the chamber.  The required termination
resistance to
prevent reflection of signals is $\unit[390]{\Omega}$.  Runs were
taken with and without termination, to see the effect of
reflected signals on PID performance.  A circuit diagram showing our
termination is in Fig.~\ref{Termination}.

Runs were taken with chambers $A$ and $B$ strung with $\unit[20]{\mu m}$
and $\unit[25]{\mu m}$ gold-plated tungsten sense wires, respectively, and
gold-plated aluminium field wires.  For some later runs, chamber $B$
was re-strung with a $\unit[30]{\mu m}$ sense wire.  The wires are
connected to the endplates by the same crimp-pins and feedthroughs
that were used in the BaBar drift chamber\cite{Aubert:2001tu}.

The gas chosen for the test was a mixture of helium and isobutane in a
$90:10$ volume ratio.  Helium was chosen because it reduces the effect
of multiple scattering compared to the more typical
argon\cite{Burchat:1991ip}.  Multiple scattering of
the charged particles is the
dominant contribution to the tracking resolution at a B-factory like
\superb.  In consideration of the rest of the \superbkerned
detector, using helium reduces the number of radiation lengths
represented by the drift chamber.  With isobutane as the quench gas, we are able to
operate the chamber with a large helium fraction, further reducing the
amount of material.
Helium also exhibits a lower drift velocity and
ionization density, which also makes it an ideal choice for cluster
counting as the incoming clusters will be less likely to overlap in
the digitized signal.  

The chambers are operated at room temperature and
atmospheric pressure. We measured the temperature and pressure during the
data taking periods, we did not use these at any level of the analysis.

\subsection{Amplifiers}\label{amplifiers}
\begin{figure}\centering
\includegraphics[width=\columnwidth,clip=true,trim = 100 110 70 100]{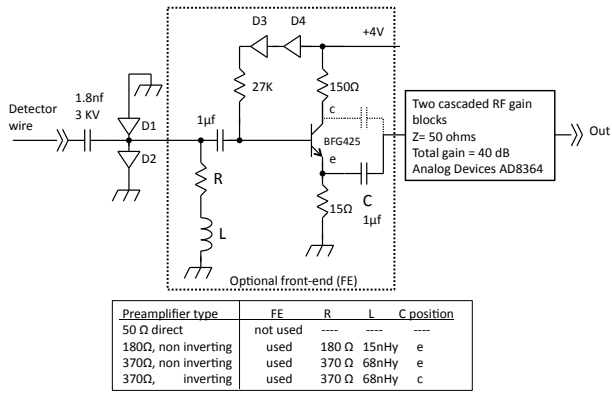}
\caption{Simplified schematic of the amplifiers used in the
  experiment.  \label{PreamSchema}}
\end{figure}

We used custom made amplifiers in order to achieve the bandwidth
required for cluster counting. The amplifiers are based on the AD8354
RF gain block from Analog Devices. These have a reasonably low power
consumption ($\sim\unit[140]{mW}$ for the whole unit) and a bandwidth of $\unit[2.7]{GHz}$. These devices have
$\unit[50]{\Omega}$ input and output impedance, and a fixed gain of
\unit[20]{dB}.  The simplest configuration that we investigated was
with two AD8354s in cascade. This provides very good bandwidth
performance, but the input impedance of $\unit[50]{\Omega}$ creates a
large mismatch with the characteristic impedance of the drift chamber
cells (around $\unit[370]{\Omega}$) and the signal to noise ratio is
not optimal.  So, an emitter follower stage was added at the input,
using a low noise RF transistor (BFG425). This was configured either
with $\unit[370]{\Omega}$ input impedance,  or with
$\unit[180]{\Omega}$, as a compromise between impedance matching and
tolerance to stray capacitance.  We also tried a configuration
with an additional low gain ($2\times$) inverting stage (with a BFG425 transistor), having
$\unit[370]{\Omega}$ input impedance.  In this case, a single AD8354
gain block was used.  The $\unit[370]{\Omega}$ configuration gave the
best overall results.  A schematic of the amplifier setup is shown in
Fig.~\ref{PreamSchema}.

In our final analysis, only the $\unit[50]{\Omega}$ and
$\unit[370]{\Omega}$ amplifiers are considered.  The data runs using
the $\unit[180]{\Omega}$ amplifiers gave signals which were of low
enough quality that a full analysis was not possible.

\subsection{Wire Voltages}\label{Wire_Voltages}
The correct voltage settings for the guard wires in
the cell were determined using the computer program
Garfield\cite{garfield}.  The guard wire voltages are chosen to make
the sensitive region of our cell behave as if it were part of an
infinite array of identical cells.  These voltages scale linearly with
the chosen sense wire voltage.

The sense wire voltage for a given chamber and amplifier was tuned to
obtain roughly equal-amplitude pulses in the final signal.  This was done
empirically by looking at the fraction of events on the oscilloscope (Sec.~\ref{DAQ})
that saturated the full voltage range.  The voltage was tuned until
this fraction was $\sim \unit[15]{\%}$.  

The resulting voltage for chamber A ($\unit[20]{\mu m}$ sense wire) using one of the
$\unit[50]{\Omega}$ amplifiers is $\unit[1700]{V}$.  The corresponding
electric field at the wire surface is calculated by Garfield to be $\unit[217]{kV/cm}$.

\subsection{Cabling}\label{cables}
For some of the runs we varied the type of signal cable used to connect the output of
the amplifiers to the data acquisition system.  We used two different
types of Sub-Miniature RG-59/U cables (models 1855A and 179DT from
Belden) and Miniature Coax (model 1282 from Belden), all with
$\unit[75]{\Omega}$ impedance.  The lengths were all $\unit[10]{m}$,
which is the distance between the amplifiers and digitizers for \superb.  From the signal-propagation
perspective, the 1855A is a better cable than the 179DT, having 
less signal attenuation ($\unit[34]{db/100m}$ versus $\unit[70]{dB/100m}$
at $\unit[1]{GHz}$).  From the perspective of mechanical integration
with the rest of the detector however, the 179DT cable would be
preferable to the 1855A, being lighter, thinner, and allowing a smaller minimum
bend radius ($\unit[25.4]{mm}$ versus $\unit[38.1]{mm}$).  

We also took data with a header connector
between the amplifier and the signal cable to simulate a connector
through the real drift chamber bulkhead.  The header connector has 20
pins with a $\unit[2.54]{mm}$ pin spacing.  Only two pins are used in the
connector to connect the ground and signal parts of an additional
$\unit[30]{cm}$ 1855A cable which is inserted in our signal cable
length using regular BNC connectors.

\subsection{Test Beam}\label{Beam}
Data were collected at the TRIUMF M11 beam\cite{M11}, which
simultaneously delivers positrons,
positive muons, positive pions at a tuneable momentum range of $100$ to
$\unit[350]{MeV/c}$.  We block residual protons from upstream using a slab of
polypropylene at the mouth of the beam pipe ($\unit[6.35]{mm}$ thick at
$\unit[210]{MeV/c}$).  We can determine
the beam populations using the time-of-flight system described in
Sec.~\ref{TOFsection}.

\begin{figure}
\resizebox{\columnwidth}{!}{%
\begin{tikzpicture}[scale=0.03]


  \fill[black] (-165,-20) rectangle (-155,-2);
  \fill[black] (-165,2) rectangle (-155,20);
  \node [align=center,above] at (-160,20) {Lead\\Collimator};

  \fill[blue] (-145,-10) rectangle (-135,10);
  \node [align=center,below] at (-127.5,-10) {Near TOF\\Scintillator};

  \fill[gray,rotate around={-10:(-20,0)}] (-25,-135) rectangle (-15,135);
  \node [left] at (-40,-70) {Chamber A};
  \fill[gray,rotate around={-10:(20,0)}] (15,-135) rectangle (25,135);
  \node [right] at (20,-70) {Chamber B};
  \node [above] at (20,135) {High Voltage End};
  \node [below] at (-20,-135) {Amplifier End};

  \fill[green] (135,-10) rectangle (140,10);
  \node [align=center,above] at (137.5,10) {Additional\\Trigger\\Scintillator};

  \fill[blue] (165,-10) rectangle (175,10);
  \node [align=center,below] at (170,-10) {Far TOF\\Scintillator};

  \node [align=center,left] at (-195,0) {Beam Pipe\\Exit};
  \draw [->,ultra thick,dashed,red] (-195,0) -- (195,0);
  \node [align=center,right] at (195,0) {Beam\\Direction};

\end{tikzpicture}}
\caption{Schematic of beam test setup at the TRIUMF M11 facility.  
  The distances in this schematic are not to scale,
  though the drift chamber proportions are correct.}
\label{Beam_Test_Schematic}
\end{figure}
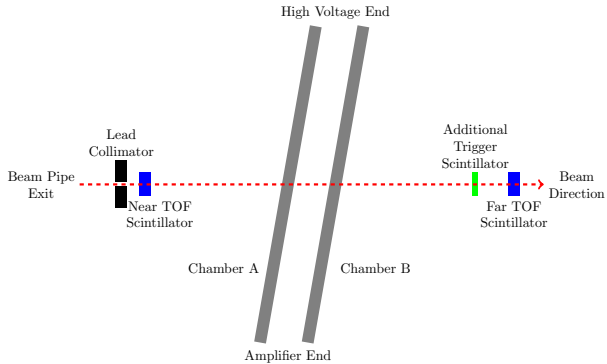

The prototypes were mounted on a rotating and moveable table, which
allowed us to take runs at different dip angles and positions along
the length of the sense wires.
A schematic of the beam test setup is in
Fig.~\ref{Beam_Test_Schematic} and a photo of the test hall is in Fig.~\ref{M11_Photo}.

Most of the data were collected at
$\unit[210]{MeV/c}$, a relatively low momentum for a high-energy
particle physics experiment.  At this momentum however the Bethe formula separation of pions and
muons is similar to the separation of pions and kaons at
$\unit[2]{GeV/c}$.  This is confirmed by our simulations at both
momenta, described in Sec.~\ref{Simulations}.  High-efficiency separation of pions and kaons at
$\unit[2]{GeV/c}$ is critical for high-precision measurements and
reconstructions at a high-energy particle experiment like \superb.

\subsection{Time of Flight}\label{TOFsection}
An external time of flight (TOF) system was used to identify the
particles independently of the prototypes.  
The beam's momentum spread is small enough that a histogram of the TOF
shows distinct peaks corresponding to the species of the particles in
the beam.
The TOF system consists of two counters $\sim\unit[4]{m}$ apart, one
upstream of the prototypes and one downstream (Fig.~\ref{Beam_Test_Schematic}).  The counters are
$\unit[12.7\times 12.7\times 220]{mm}$ BC-404 scintillators each read out by
two Burle 8501-1 64-channel micro-channel plates (MCPs), one on each end of
the scintillator block.  The scintillators are roughly
the same size as the beam spot.  The MCPs have $\unit[25]{\mu m}$
pores.  Each of the 64 channels in the MCPs have an active region of
$\unit[6\times 6]{mm}$.  We gang together four of the channels to form
one combined signal.  This signal from
each MCP goes to an Ortec 935 constant-fraction discriminator (CFD)
with no pulse height correction applied.  Each is then delayed by a given
time in order to separate the pulses, then they are combined and recorded in a
single channel of our oscilloscope.  

The signals from the MCPs are used as part of the trigger.  Additional
signals from photomultiplier tubes are used and are described in
Sec.~\ref{trigger}.

The unscaled TOF is obtained by determining the arrival time of each pulse
from the MCPs as recorded by the oscilloscope.  The first two pulses
are from the two ends of the upstream counter, while the following two are
from the downstream counter.  These pairs are averaged, then the
difference is taken.  There are arbitrary delays 
associated with the MCP signals, so the TOF quantity is scaled to be
physically meaningful.  A run is chosen and a histogram of the TOF
quantity is made, where the positrons, muons, and pions are clearly
resolved as Gaussian peaks.  We fit the positron peak with a Gaussian distribution.  The beam momentum is high enough that the positrons may be
treated as moving at the speed of light, and the actual distance between the two counters is well-measured. 

\begin{figure}\centering
\includegraphics[width=\columnwidth,clip=true,trim = 0 0 40 30]{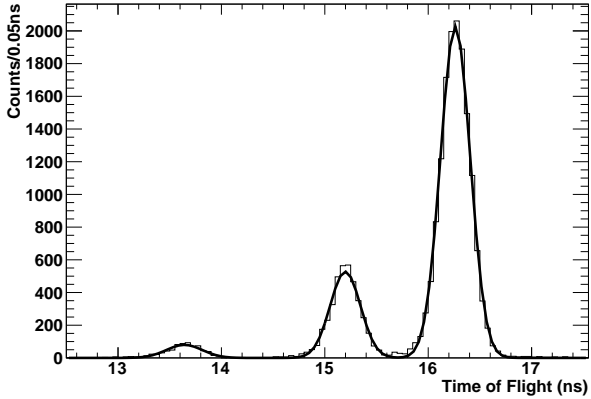}
\caption{Time-of-flight histogram for a run at $\unit[210]{MeV/c}$
  beam momentum.  The three peaks correspond to positrons, muons, and
  pions, in increasing TOF order.  The fit is to the sum of three
  Gaussians.}
\label{TOF}
\end{figure}

We were able to achieve TOF resolutions of $\sim\unit[60]{ps}$ per
MCP.  For a $\unit[210]{MeV/c}$ beam (Fig.~\ref{TOF}) the separations
of the Gaussian peaks are greater than the $\unit[3]{\sigma}$ ranges
used to identify particles in our track composition process described in Sec.~\ref{bootstrapping}.  A sample
trace of the actual TOF signal is shown in Fig.~\ref{sampletraces}, where the first
four pulses are from the MCPs.

We fit the TOF distribution with the sum of three Gaussians and count how many
particles are within $\unit[3]{\sigma}$ of each peak.  For the run
shown in Fig.~\ref{TOF}, we find that of all the physical triggers 
$3.8\%$ are positrons, $20.5\%$ are muons, and $75.7\%$ are pions.
 
\subsection{Trigger}\label{trigger}
The TOF signals are also used as part of the trigger system for the
oscilloscope.  It was noted that with only the upstream and downstream
counters, many events contained no signals in the drift
chambers (i.e.\ oscilloscope traces with just normal noise, no
clusters).  
In addition, the TOF histogram showed six peaks, though only three were
expected.  The six peaks appeared to be in two similar groups, shifted
in TOF value.  The conjectured origin of the
higher-TOF valued population was beam particles passing through the
upstream counter but angled downwards, scattering off of
the metal table, then passing through the downstream counter,
bypassing the chambers entirely and taking a longer path.  

\begin{figure}\centering
\includegraphics[width=\columnwidth,clip=true,trim = 50 10 200 0]{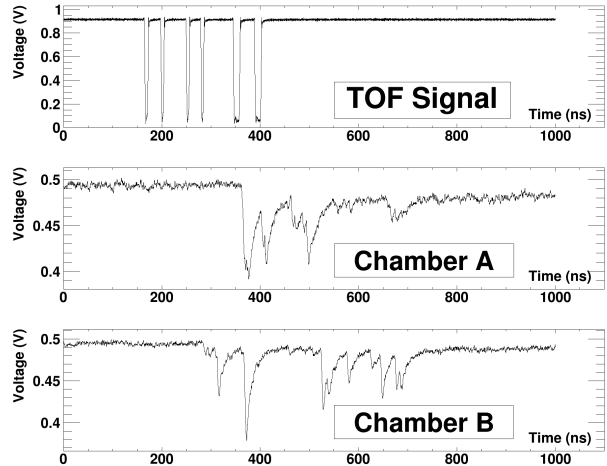}
\caption{Oscilloscope traces for a run at $\unit[210]{MeV/c}$
  beam momentum.  The first is the TOF signal, with four initial
  pulses from the TOF MCPs, and two additional
  pulses from the extra trigger PMT.  The TOF value identifies
  this particle as a pion.  The second and third traces are
  from prototypes $A$ and $B$, respectively.  The cluster structure is
  clearly evident in these signals.}
\label{sampletraces}
\end{figure}

A third scintillator strip $\unit[3]{mm}$ thick was
placed between the prototypes and the downstream counter (Fig.~\ref{Beam_Test_Schematic}), instrumented
with photomultiplier tubes.  The coincidence of the three (upstream,
downstream, strip) was required for a physical trigger.  This
additional requirement removed the extraneous TOF population and
many of the events with no drift chamber signals.  Part of the trigger
signal can be seen in Fig.~\ref{sampletraces} in the upper trace.  The
third scintillator was not digitized and thus is not visible in the figure.

The coincidence rate is $\sim\unit[30]{Hz}$, while the signal rate on
the upstream counter without requiring coincidences ranges from a few
\unit{kHz} to tens of \unit{kHz}, depending on beam line settings.  We also
introduced an asynchronous trigger based on a pulse generator whose
frequency was tuned to $\sim 15\%$ of the total trigger rate.  These
asynchronous triggers are uncorrelated with real beam events.  They
provide a sample of empty events for monitoring and measuring baseline
voltages and noise levels during the run.

\subsection{Data Acquisition System}\label{DAQ}
Our data acquisition system consisted of a LeCroy WavePro 740Zi, an
oscilloscope with 4GHz bandwidth.  Data were written to an external
USB hard disk in a proprietary binary format and then converted into
ROOT\cite{ROOT} files for analysis.  The oscilloscope writes one file per active
channel per trigger.  We used one channel for the time-of-flight system
and one channel for each prototype sense wire, meaning we had three
small files written per trigger.  Each channel read $20002$ samples
with $\unit[50]{ps}$ spacing, for a trace duration of
$\sim\unit[1]{\mu s}$.  The biggest
bottleneck was the filesystem (Microsoft NTFS), which does not perform
well with directories having tens of thousands of files.  The overall
rate of events written to disk was $\sim\unit[12]{Hz}$.

We used the MIDAS\cite{MIDAS} data acquisition system to automatically record temperature and atmospheric
pressure as well as the current in a small monitoring chamber.  The
monitoring chamber was connected in series with the primary chambers on
the gas line, and was exposed to an \ce{^{55}Fe} source.  The
monitoring chamber wire voltages were held fixed, allowing us to monitor
the gas and environmental conditions by tracking changes in the gas gain.

\section{Simulations}\label{Simulations}
We used a gaseous ionization detector simulation package called
Garfield\cite{garfield} to simulate tracks through our prototypes.
We did not simulate the electronics chain and the data acquisition
system, but we are able to get predicted charge depositions and
cluster counts for our specific gas mixture and wire configuration.

\begin{figure}\centering
\includegraphics[width=\columnwidth,clip=true,trim = 0 0 40 0]{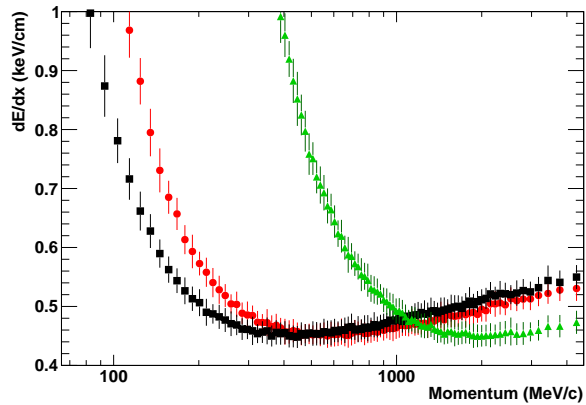}
\caption{Garfield simulation of the energy loss by a charged particle
  crossing 40 cells of a 90:10 mixture of helium and isobutane.  The
  black squares, red circles,
  and green triangles represent muons, pions, and kaons, respectively.
  The marker position is the \unit[70]{\%} truncated mean energy loss, while the
  vertical error bar on each marker is the RMS of the truncated mean.}
\label{Garfield_dEdx}
\end{figure}

The charge deposition is not reported directly, but is proportional to
the energy lost by charged particles passing through the gas.  It
is plotted in Fig.~\ref{Garfield_dEdx} for muons, pions, and
kaons.  The momentum scale is chosen to illustrate the fact that
the difference in energy loss between pions and muons at
$\sim\unit[200]{MeV/c}$ is similar to that between pions and kaons at
$\sim\unit[2]{GeV/c}$ (Sec.~\ref{Beam}).

\begin{figure}
  \centering
  \begin{tikzpicture}
    \node[anchor=south west,inner sep=0] (image) at (0,0){
      \includegraphics[width=\columnwidth,clip=true,trim = 0 0 40 0]{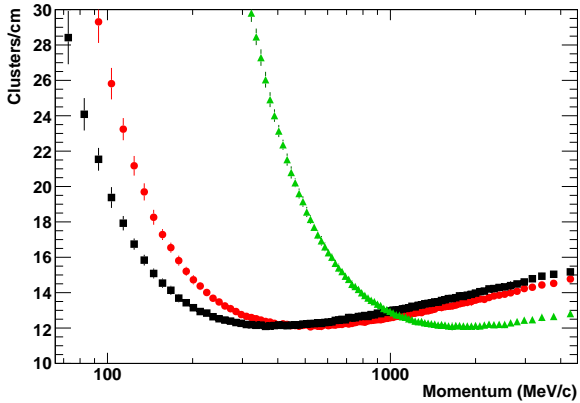}};
    \begin{scope}[x={(image.south east)},y={(image.north west)}]
    \end{scope}
    
  \end{tikzpicture}
  \caption{Garfield simulation of charge clusters produced by a charged particle
    crossing a 90:10 mixture of helium and isobutane.  The black
    squares, red circles,
    and green triangles represent muons, pions, and kaons, respectively.
    The marker position is the average number of clusters, while the
    vertical error bar on each marker is the RMS.}
  \label{Garfield_Clusters}
\end{figure}

The number of primary ionizations is reported directly by the
simulation software and can be treated as a ``true'' number of
clusters.  It does not depend on the choice of electronics,
algorithms, and it does not count $\delta$-rays (Sec.~\ref{CC_PID}).  The
distribution of primary ionizations for muons, pions and kaons is
shown in Fig.~\ref{Garfield_Clusters} and also shows the similarity
between muon-pion separation at our beam momentum and pion-kaon
separation at higher momenta.  It is also important to point out that
the absolute number of clusters for muons and pions at
$\unit[210]{MeV/c}$ approximately mirrors that of pions and kaons at
$\unit[2]{GeV/c}$, not just the difference.  The absolute value is
important because it is related to our ability to actually resolve the clusters.

\section{Beam Test Data}
The data were taken during August and September 2012.  Approximately
200 runs of 30000 events were acquired.  A run is a contiguous
data-collection period during which no setup parameters are changed.
On average, $\unit[15]{\%}$ of the events were from
asynchronous triggers and $\unit[10]{\%}$ of the physical triggers did not leave
signals in the prototypes.

Various parameters were changed from run to run.  These were: the
sense wire voltages, amplifiers, signal cable types, beam momentum,
angle of incidence of the beam with the chamber, beam position along
the sense wire length and presence of a proper termination resistor on
the sense wire.   In the end, many runs turned out to be recorded
using unsuccessful amplifier prototypes and could not be used for a
detailed analysis.  This analysis uses 20 runs, for a total of
633050 recorded events.

\section{Analysis}
The analysis of the test-beam data is performed in two steps, both of
which are done offline (after the data for that run has been fully collected).  The
first step involves analyzing the signals (voltage as a function of time)
from the three oscilloscope channels.  The first channel is connected to the
time-of-flight (TOF) system, with voltage pulses corresponding to a
particle crossing the scintillators before and after the drift
chambers.
The second and third oscilloscope channels are connected to the
amplifiers on the sense wires of the two drift chambers.

The second step of analysis involves constructing multi-cell
``tracks'' from the single-cell events using a composition process.
Single-cell events are taken from the same run, same chamber, and having a TOF
consistent with the same particle type.  Forty of these are used to
build up a track as if it were traversing a full \superb-size drift
chamber (Sec.~\ref{bootstrapping}).

\subsection{Single-Cell Information}
This section describes in detail the first stage of analysis in which
we deal with single-cell events.  The time-of-flight is measured, the
signal is adjusted for baseline drift and basic quality controls are
imposed.  In this stage we also perform the charge integration and use
cluster-counting algorithms to count clusters on the drift chamber signals.
\subsubsection{Time of Flight}
The time-of-flight is determined by applying a simple
threshold-over-baseline algorithm to the oscilloscope trace from the
channel connected to our scintillator MCPs and PMTs.  A valid TOF
signal consists of four identified pulses, while an asynchronous
trigger has zero pulses.
Events with one, two, or three TOF pulses are rejected, and represent the small fraction of
events from asynchronous triggers with a pulse in one of the TOF counters.

\subsubsection{Baselining and Signal Confirmation}\label{baselining}
The baseline voltage for each drift chamber is simply the average
voltage of
the entire signal from the previous asynchronous trigger.  The RMS
deviation from this baseline is also measured.
The mean of these RMS deviations is $\sim\unit[2]{mV}$.  Signals from
physical triggers have amplitudes on the order of hundreds of $\unit{mV}$ above the baseline.

The real particle events are tested for the presence of an actual
signal by a threshold algorithm, where the baseline and threshold
levels are determined by the previous asynchronous trigger
measurements.  Real particle events that have no signal in the
chambers are rejected.  These are from events where a real particle
crossed the scintillators, but either missed one or both drift
chambers, or did not interact within them. 

\subsubsection{Charge Integration}
\begin{figure}\centering
\includegraphics[width=\columnwidth,clip=true,trim = 0 0 40 30]{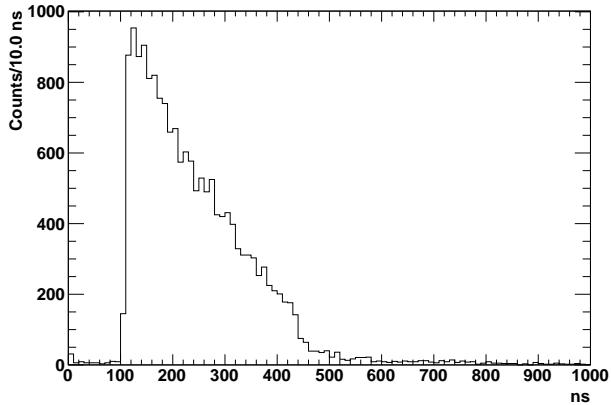}
\caption{Time at which the charge integration begins in Chamber $A$ for
  a run at $\unit[210]{MeV/c}$.}
\label{startintegral}
\end{figure}

\begin{figure}\centering
\includegraphics[width=\columnwidth,clip=true,trim = 0 0 40 30]{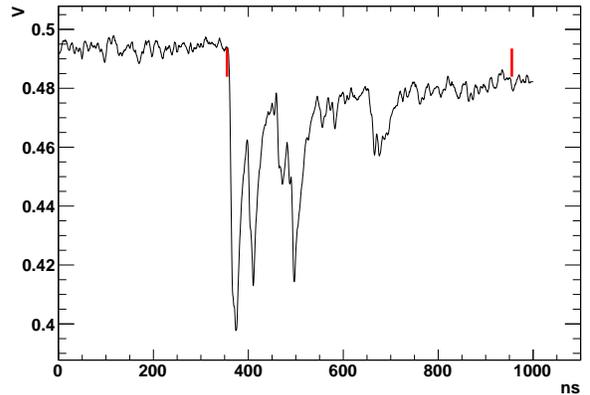}
\caption{Sample event in chamber $A$ showing \unit[600]{ns} integration time.}
\label{integralduration}
\end{figure}

A charge integration is performed for the remaining asynchronous and
physical events,
starting at the time of the threshold crossing mentioned in \ref{baselining} (or at
an arbitrarily chosen time for asynchronous events), integrating for a
fixed duration.  The distribution of start times for a sample run is
shown in Fig.~\ref{startintegral}.  If the duration is too short, then some pulses may be missing or the tail of the last pulse
may be clipped.  If the duration is too long, then unnecessary noise
is also integrated, reducing the resolving power of the charge
measurement.  Different equipment combinations give different pulse
tail decay times, so the duration must be optimized empirically.
A typical optimal value is $\sim\unit[600]{ns}$, as shown in
Fig.~\ref{integralduration}.  The optimization of the integration time
is described in Sec.~\ref{ChargeIntegration}.

\begin{figure}\centering
\includegraphics[width=\columnwidth,clip=true,trim = 0 0 40 30]{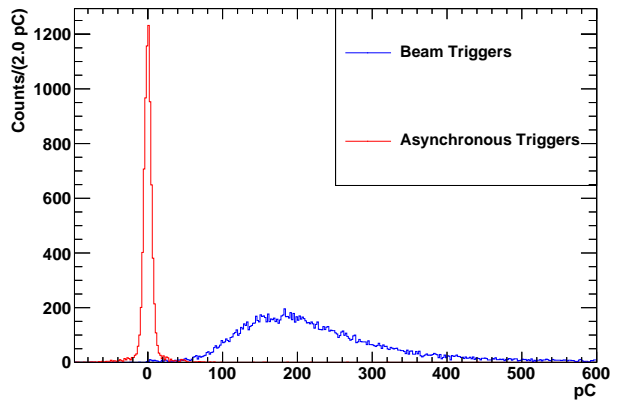}
\caption{Baseline-subtracted charge distributions as identified by the
time-of-flight system.  The sharp peak on the left is from
asynchronous triggers (with no particles in the prototypes), while the broader peak in the middle is from
physical triggers with all particle species combined.}
\label{cellcharge}
\end{figure}

\begin{figure}\centering
\includegraphics[width=\columnwidth,clip=true,trim = 0 0 0 10]{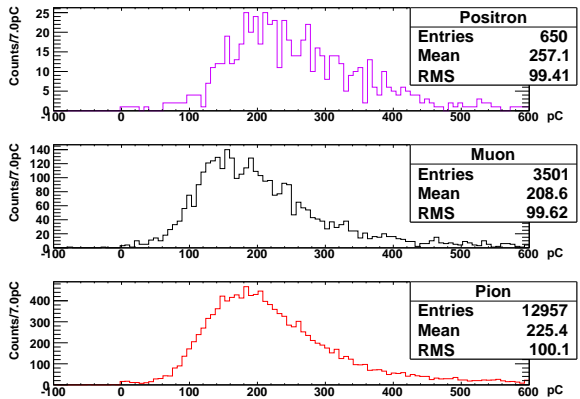}
\caption{Baseline-subtracted charge distributions for each particle
  species at $\unit[210]{MeV/c}$.  Note that the sample mean and RMS values
  indicated in the figure are not representative of the underlying
  distribution since it does not have well-defined moments.}
\label{CellChargePerSpecies}
\end{figure}

From the integrated charge we subtract a pedestal calculated from the
previous asynchronous trigger.  This pedestal is a charge integration
with the same integration time, but a fixed starting time.
The result is a baseline-subtracted charge, which should have a
smaller systematic error than the raw charge integral.  The
distribution of integrated charges for physical triggers and
asynchronous triggers is shown in Fig.~\ref{cellcharge}.  The physical
triggers are shown separately for each species in Fig.~\ref{CellChargePerSpecies}.

\subsubsection{Cluster Counting}
Cluster-counting algorithms can vary in complexity,
efficiency, and in their rate of reporting fake clusters.  Here we briefly describe the
various algorithms, but precise
definitions can be found in Appendix \ref{CC_appendix}.  

\begin{figure}\centering
\includegraphics[width=\columnwidth,clip=true,trim = 0 0 40 10]{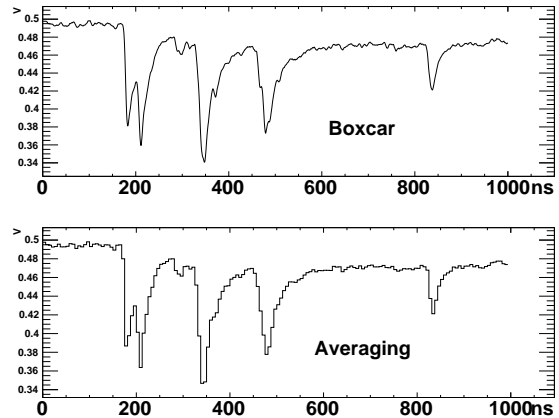}
\caption{The two smoothing algorithms used, each smoothing over 125
  frames of $\unit[50]{ps}$ width, for a smoothing width of
  $\unit[6.25]{ns}$.  This event is the same as shown in
  Fig.~\ref{integralduration}}
\label{fig:smoothings}
\end{figure}

The algorithms involve two forms of smoothing of the oscilloscope
traces (Fig.~\ref{fig:smoothings}).  The first is a
``boxcar smoothing'' where each sample is replaced with the average of
itself and the $n-1$ previous samples.  The second is a true averaging
procedure, where the number of points in a trace is reduced and each
point is the average of $n$ points.

\begin{figure}\centering
\includegraphics[width=\columnwidth,clip=true,trim = 50 0 150 10]{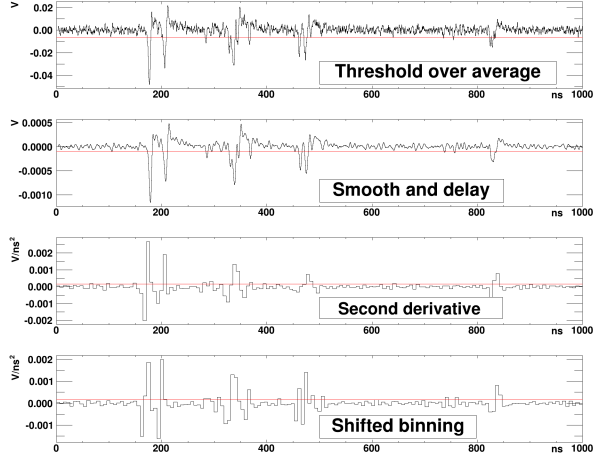}
\caption{Illustration of the quantity on which a threshold is applied
  in the various cluster-counting algorithms.  Each uses a set of
  parameters (smoothing width, threshold level) that were optimized
  for this run.  The threshold level is indicated by the red
  horizontal line.  The last image is the same as the second derivative,
  but with the binning shifted, to show that some clusters can be
  hidden by the binning (e.g.\ around \unit[480]{ns}).  This event is
  the same as shown in Figs.~\ref{integralduration}
  and~\ref{fig:smoothings}}
\label{CC_Cut_Quantities}
\end{figure}

All of the algorithms involve some kind of transformation of the
smoothed signal, and a threshold-crossing criterion.  The transformed
signals for the various algorithms are shown in Fig.~\ref{CC_Cut_Quantities}.
One of the most basic cluster-counting algorithms is the ``Threshold
above Average''.  It subtracts the non-smoothed signal at time $t$ from
the boxcar-smoothed signal at time $t-1$, then applies a threshold.  

A more general algorithm (of which the previous is a special case) is
the ``Smooth and Delay'' algorithm.  It involves smoothing two copies
of the signal by different amounts, delaying one of the copies by a
certain number of frames, then taking the difference and applying a
threshold.  This algorithm has four parameters, and is thus more difficult to optimize.

\begin{figure}\centering
\includegraphics[width=\columnwidth,clip=true,trim = 20 0 10 10]{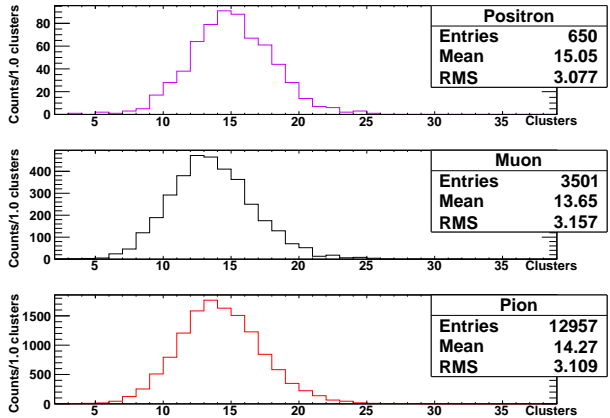}
\caption{Number of clusters found for each species as identified by
  the TOF system.  This is for a \unit[210]{MeV/c} run using the
  second-derivative algorithm.}
\label{CellClustersPerSpecies}
\end{figure}

The two algorithms above essentially implement a first-derivative method.  We
also implemented a second-derivative method.  This one uses the true
averaging procedure rather than the ``boxcar smoothing''.  The first
derivative is first calculated by taking the difference between
consecutive smoothed samples.  The second derivative is then
calculated by taking the difference between consecutive first
derivative values.  Each time, we divide by the time interval
represented by a sample, to keep the units consistent.  The number of clusters counted using
the second derivative is shown for each particle species in Fig.~\ref{CellClustersPerSpecies}.

All of the threshold algorithms in principle trigger on the leading
edge of cluster signals.  However it is noticeable that real cluster
pulses have a very sharp leading edge (approximately $\unit[3]{ns}$) and a slower decaying trailing
edge (approximately $\unit[100]{ns}$).  Fake clusters are more symmetric, returning to the
baseline voltage faster than the signal from a real cluster.  Thus an
algorithm was devised that takes cluster candidates from the above
algorithms, but requires the pulse to last a minimum duration in order
to be confirmed.  Pulses that return to baseline too quickly are
discarded as fake clusters.  This ``timeout booster'' allows the use
of smaller thresholds, which while increasing the efficiency of finding real
clusters also admit more fakes.  The timeout criterion removes most of
the fakes but keeps the real clusters.

As mentioned before, each of the cluster-counting algorithms can return not only the number
of clusters, but the actual time at which each cluster was found.  We
investigated the use of this information, in the form of an average
time separation between clusters in each cell.  

\subsection{Track Composition}
\label{bootstrapping}
The prototypes have only a single cell.  The traditional method of
identifying particles using the truncated mean requires many cells
forming a track.  Thus we construct tracks from the single-cell events.

To compose a track for a given species of particle, we
select (with replacement) random single-cell events that have been identified with the
time-of-flight information.  We positively identify particles with TOF
values within 3 standard deviations of the central values of the
three Gaussian peaks corresponding to the particle species.
For a typical run with e.g.\ 3500 single muon events, the number of possible
muon tracks is astronomical ($\sim 10^{94}$), and the likelihood of a given
track being composed of multiple copies of the same single-cell event
is low ($\sim 1\%$).  We also form empty tracks by combining the
signals from asynchronous events.

The information from each event is
combined to form the track information.  The track information is the
particle species, total number of clusters found per \unit{cm} of
track, and the truncated mean of the charge integrals from each cell.
The truncated mean is performed by sorting the list of charge
integrals and taking $\unit[70]{\%}$ of the values starting from the
beginning of the list.  The value of $\unit[70]{\%}$ was roughly
optimized to give better separation, for comparison $\unit[80]{\%}$
was used in BaBar\cite{Aubert:2001tu}.  The \superbkerned drift
chamber design has 40 layers.  Thus we use 40 events from our
single-cell prototypes events to create a composed track.  
The $\unit[70]{\%}$ truncated mean was thus done by rejecting the
largest 12 integrated charge values from the cells.

\begin{figure}\centering
\includegraphics[width=\columnwidth,clip=true,trim = 0 0 40 30]{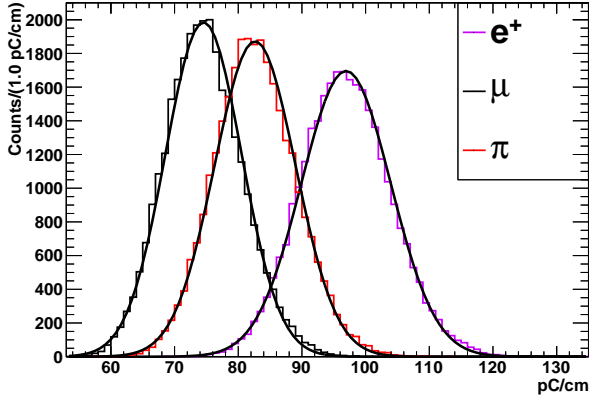}
\caption{Truncated mean of charges ($dE/dx$) in composed tracks.  This is using
  events from the same run as Fig.~\ref{CellChargePerSpecies}.  The
  three peaks from left to right are from muons, pions, and positrons,
  respectively.  Here the particle populations are equal, as we
  compose an equal number of tracks for each species.}
\label{TrackChargePerSpecies}
\end{figure}

\begin{figure}\centering
\includegraphics[width=\columnwidth,clip=true,trim = 0 0 40 30]{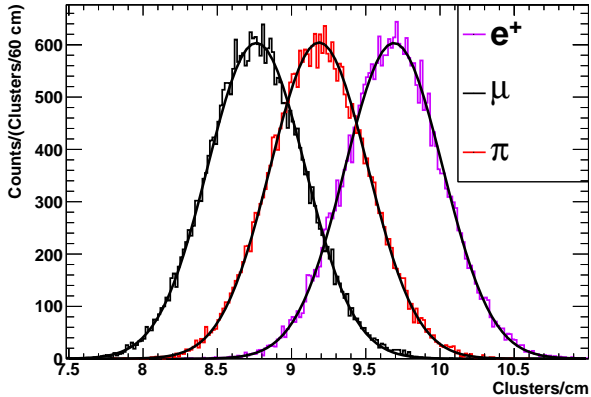}
\caption{Number of clusters per $\unit{cm}$ in composed tracks.  This is using
  events from the same run as Fig.~\ref{CellClustersPerSpecies} and using the
  second-derivative algorithm.  The
  three peaks from left to right are from muons, pions, and positrons,
  respectively.  Here the particle populations are equal, as we
  compose an equal number of tracks for each species.}
\label{TrackClustersPerSpecies}
\end{figure}

In the case of tracks formed from asynchronous
events, the list is not sorted, since these values are already Gaussian, but still the same
fraction of values is discarded.  The distribution of truncated mean
charge and clusters for the composed tracks is shown in
Figs.~\ref{TrackChargePerSpecies} and \ref{TrackClustersPerSpecies},
respectively.  

We also form the track-wise average time separation between clusters by
doing a weighed average of the cell-wise average cluster separation
for the events in the track.  The weights are the number of clusters in the cells.

It is worth noting that the relative separations of the muon and pion
peaks shown in Figs.~\ref{TrackChargePerSpecies} and
\ref{TrackClustersPerSpecies} are very different.  For the truncated
mean of the integrated charges, the relative separation between the
peaks (difference in the location of the peaks, divided by the
average of the two) is $\sim\unit[10]{\%}$, while for the cluster
counting it is $\sim\unit[5]{\%}$.  Na\"ively this should mean that the
cluster counting technique is less effective.  However because the widths
of these peaks is also very different, the two techniques turn out to
be of comparable power (Fig.~\ref{R_and_dEdx}).

\subsection{Combined Likelihood Ratio}
In order to combine the information from the truncated mean and the
cluster count, we form likelihoods based on fits
to the two quantities.  These quantities are
reasonably Gaussian (for non-empty tracks), so we fit them with
Gaussian distributions $G_{s,k}$, for particle species $s$ and
measured quantity $k$.  For a given track, the likelihood of the
track coming from a particle $s$ is found by evaluating
the product of the fitted
distribution functions for both $k$s at the measured values.  Thus if
the measured truncated mean charge for a track is $q$ and the clusters
per $\unit{cm}$ of track are $n$, the combined likelihood is

\begin{equation}
L_{s}(q,n) = G_{s,\text{charge}}(q) \times G_{s,\text{clusters}}(n).
\end{equation}

This combined likelihood ignores any correlation between the two
quantities.  The correlation is indeed non-zero but is somewhat weak
($\sim 0.3$).  Possibly combined likelihood models which make use of
the correlation would be more effective, but we did not investigate this.

As mentioned in Sec.~\ref{Beam}, the ability to identify muons and
pions at $\sim\unit[210]{MeV/c}$ is our proxy variable for the
performance of the prototypes.  Thus we
form a ratio of the combined likelihoods of being a muon and pion:

\begin{equation}
R(q,n) = \frac{L_{\mu}(q,n)}{L_{\mu}(q,n)+L_{\pi}(q,n)}. \label{R}
\end{equation}

\begin{figure}\centering
\includegraphics[width=\columnwidth,clip=true,trim = 0 0 40 30]{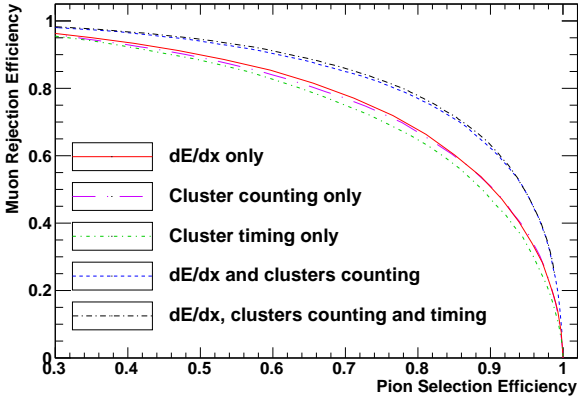}
\caption{Efficiency graph for a run at \unit[210]{MeV/c}, the same run
  as earlier figures.  The cluster
  counting is done using the second-derivative algorithm.  The upper
  two curves nearly coincide and are the efficiencies when cutting on the combined
  likelihood ratios.  One combines the truncated mean, cluster count,
  and cluster separation, the other only truncated mean and cluster
  count.  The lower three curves are the efficiencies when one cuts directly
  on the truncated mean, cluster count, or cluster separation
  quantities.}
\label{R_and_dEdx}
\end{figure}

This quantity's distribution is peaked at 0 for real pions and at 1
for real muons.  A cut can be made that maximizes the separation according to
some figure of merit.  A typical way to demonstrate the performance is
by making a rejection-selection efficiency plot.
Consider the fraction of real pions that would also be identified as pions by
the cut on $R$, and the fraction of real muons that would not be
identified (that is, rejected) as pions by the cut on $R$.  We can
thus make a parametric plot of muon rejection efficiency on the
vertical axis
and pion selection efficiency on the horizontal axis, with the parameter being
the chosen $R$ cut value (Fig.~\ref{R_and_dEdx}).  Similar efficiency graphs can be made for
cuts directly on the physical quantities of charge and cluster
counts.

\subsection{Figures of Merit}\label{FOM}
The efficiency graphs are a good way to represent the performance of a
particular setup, but they are two-dimensional and difficult to
include in summaries.  Thus we construct figures of merit in order to
quantify the performance of an equipment choice or algorithm.  A
convenient method is to set a given background rejection level and
state the corresponding signal efficiency.
In the muon rejection and pion selection plot, one may thus find the muon
rejection efficiency corresponding to $\unit[90]{\%}$ pion selection efficiency, or
vice-versa.  These figures of merit are easy to interpret
physically and correspond to how detector performance is typically
quantified in past experiments.

An alternative figure of merit turns out to better
differentiate between algorithm parameter choices, but has a much less
intuitive physical meaning.  It is the maximum excursion on the muon
rejection and pion efficiency plot from the origin of the graph.  The
curves on the graph approach $(0,1)$ and $(1,0)$ in the limits of $R$
cut values of 0 and 1 respectively, but the curves can lie above that
inscribed by a circle of unit radius.  The length of the longest
straight line joining $(0,0)$ and the efficiency curve is taken as the
figure of merit.  In certain cases the performance is bad enough that the
lines lie below that inscribed by a circle, in this case the alternative
figure of merit is not meaningful, as it is identically 1.

All three figures of merit can be shown to be equivalent, in the sense
that local maxima and minima lie in the same regions of parameter
space.  The maximum-excursion-from-origin figure gives better
separation for those runs where it is meaningful (the majority).  It
is used for the optimization of algorithms, but the results are
presented using the more intuitive figure of merit of pion selection efficiency at $\unit[90]{\%}$ muon rejection.

\section{Results}
In this section we present the results of varying the cluster-counting
algorithms, gas gain, various chamber positions, and other equipment
choices.  

\subsection{Charge Integration}\label{ChargeIntegration}
\begin{figure}\centering
\includegraphics[width=\columnwidth,clip=true,trim = 0 0 40 30]{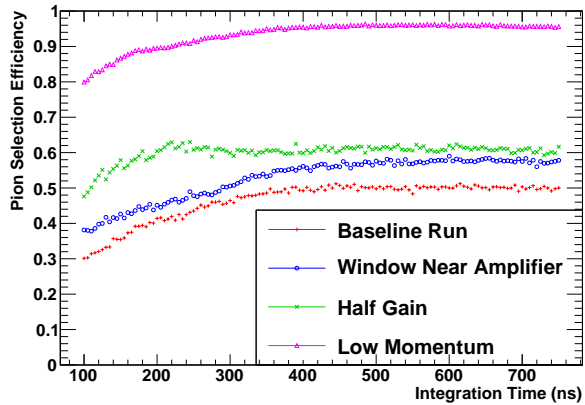}
\caption{Pion selection efficiency using $dE/dx$ only for several runs
  as a function of charge integration time.  The baseline run is at
  a window $\unit[1883]{mm}$ from the amplifier, $\unit[210]{MeV/c}$ and using nominal gain as calculated
  by simulations.  The low-momentum run is at $\unit[140]{MeV/c}$.
  All the runs use the extra termination resistor.}
\label{ItimeVariation}
\end{figure}

The time over which to integrate a signal in order to capture the
charge deposition on the wire was determined empirically.  In
principle the optimal value varies from run to run depending on gas gain, dip
angle of the beam, and window position, but we wish to compare runs at
different settings.  Thus we look at the figure of merit for many
different runs and choose a suitable compromise (Fig.~\ref{ItimeVariation}).  As it turns out, the
performance does not vary strongly as a function of integration time
once the time is suitably long.  We choose an integration time
of $\unit[600]{ns}$ for the rest of the study.

\subsection{Cluster Counting}\label{CC_Results}
\begin{figure}\centering
\includegraphics[width=\columnwidth,clip=true,trim = 0 0 0 0]{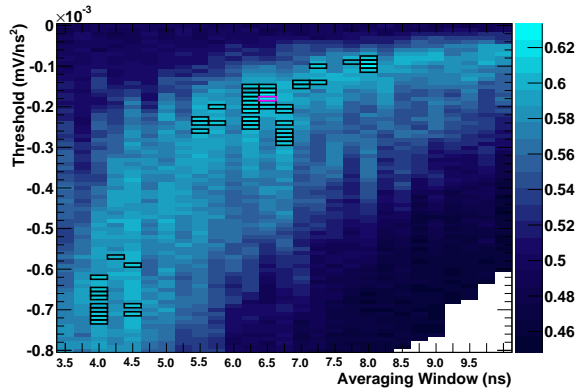}
\caption{Example performance ``heat map'' for the second-derivative
  algorithm using a run at \unit[210]{MeV/c}.  In this case, the
  optimal parameters are an averaging window of $\sim\unit[6.5]{ns}$
  and a threshold of $\sim\unit[-0.16]{mV}$.  The black rectangles are
  the highest-performance regions, the magenta is the single best.}
\label{CC3_Optimization}
\end{figure}

The various cluster-counting algorithms have parameters that must be
tuned empirically.  By iterating this procedure many times using the same
run, a ``map'' of the figure of merit can
be created in the algorithm parameter space, the maxima of which are
optimal values for the algorithms (Fig.~\ref{CC3_Optimization}).  While the figure of merit includes
the PID performance from $dE/dx$ and cluster counting, the $dE/dx$
contribution is essentially constant even with the randomness introduced by
the track composition process.  

The optimal parameters vary from algorithm to algorithm and depend
on the run used to optimize the parameters.  In an operational
experiment, only one set of parameters can be chosen, so some
compromise will be necessary.  Nevertheless, to compare the algorithms
themselves, we may compare the performance of each algorithm when
optimized on the same data run.  

The chosen run has the following parameters: $10$ degree dip angle,
window $\unit[1883]{mm}$ from the amplifier, and $\unit[20]{\mu m}$ sense
wire.  A $\unit[370]{\Omega}$ inverting amplifier and 1855A
Sub-miniature RG59/U signal cable with no extra connector are used.
The beam momentum was $\unit[210]{MeV/c}$.  A total of 30784 triggers
were recorded of which 7720 are asynchronous, and 680, 3649, and 13579
are positively identified as  positrons, muons, and pions
respectively.  The remainder have TOF values more than
$\unit[3]{\sigma}$ away from the peaks or have no signal in the chamber.

The dip angle of 10 degrees was chosen rather than 0 in order to avoid
space charge effects. The avalanches produced on the wire from the passage of a particle at zero
degrees occur all in the same gas volume near the wire.  This can
affect the overall results and essentially makes 0 degrees a
``special'' angle.  In an operating $e^+e^-$ collider experiment the
fraction of tracks suffering from space charge effects is negligible.

We explored a variety of algorithms, which are described in detail in
Appendix \ref{CC_appendix}.  
Each has some kind of threshold as one of
the parameters, and some smoothing or averaging duration.  A common
feature is that the optimal smoothing or averaging duration is
$\sim\unit[5]{ns}$, which indicates that extremely high sampling
rate and bandwidth are not necessary to improve PID with cluster counting.  The
smoothing times correspond to Nyquist frequencies of
$\sim\unit[100]{MHz}$.  Our amplifiers have much higher bandwidth than
this (Sec.~\ref{amplifiers}), so using amplifiers with smaller bandwidth but
better signal-to-noise ratios should improve overall performance.

\begin{table}\small
\centering
    \caption{Summary of optimal parameters for the various
      cluster-counting algorithms.  The algorithms labelled $A$, $B$,
      $C$, $D$, and $E$ are ``Signal above
      Average'', ``Smooth and Delay'', ``Signal above Average with
      Timeout'', ``Second Derivative'', and ``Second Derivative (Two
      Passes)'', respectively.  The threshold is given with the
      appropriate units for that algorithm, and $\tau$ is the smoothing or averaging time in
      nanoseconds.  Algorithm $B$ has in principle two
      smoothing times, but the optimal value has them equal.  The
      additional parameter $\Delta_t$ for the algorithms $B$ and $C$ are the
      delay and the timeout, respectively.  The figure of merit
      $\epsilon_{\pi}$ is the pion selection efficiency for
      $\unit[90]{\%}$ muon rejection.  \label{CC_table}}
    \begin{tabu} to \columnwidth {X[1.5]|X[1.5]|X[0.7]|X|X[0.7]}
      Algorithm & Threshold & $\tau (\unit{ns})$ & $\Delta_t (\unit{ns})$ & $\epsilon_{\pi}$ \\
      \hline $A$ & \unit[-6.5]{mV} & 3.5 & & 0.62 \\
      \hline $B$ & \unit[-0.1]{mV} & 2.75 & 3.75 & 0.64 \\
      \hline $C$ & \unit[-3.0]{mV} & 3.5 & 4.25 & 0.62 \\
      \hline $D$ & $\unit[0.16]{mV/ns^2}$& 6.5 & & 0.64 \\
      \hline $E$ & $\unit[0.15]{mV/ns^2}$ & 6.25 & & 0.64\\
    \end{tabu}
\end{table}

In Table~\ref{CC_table}, the figure of merit is the pion selection
efficiency for $\unit[90]{\%}$ muon rejection.  Here and in later
plots, it is difficult to give a good estimate of the systematic uncertainty as
many factors were not taken into account.  For example the temperature
of the gas in the chamber plays no role in our calculations, though
the temperature did change during the data taking period.  The
track composition process involves drawing random numbers, so a
contribution to the uncertainty from this can be estimated by
composing multiple sets of tracks and seeing the distribution of
results.  Running the code 100 times yields an RMS deviation from the
mean of $\sim 0.05$.  The mean is what is reported in Table~\ref{CC_table}.

In the table, only algorithm $C$ uses the ``Timeout Booster''
technique.  We also tried applying the technique to the other algorithms,
but it was noticed that if the algorithm already has reasonable performance, the
improvement from the timeout is negligible.  Indeed the optimal
timeout duration for the ``Smooth and Delay'' algorithm is zero,
yielding the same performance as the bare algorithm.

Overall the best algorithm is the two-pass second derivative algorithm, but it is only
marginally better than the other algorithms.  The difference is less
than the typical variation due to the track composition process.

It is fortuitous that even the simple algorithms have good performance,
as they are reasonable to implement using a
field-programmable gate array (FPGA) or even analog hardware.  

In some sections that follow, the PID performance with optimized
cluster counting refers to the use of a cluster-counting algorithm
where the parameters were chosen to give the best figure of merit for
that run.  The optimal parameters vary from run to run, so in each
case, we also run the algorithm on a given run using parameters that
were optimal for a set of other runs.  The other runs each vary in
only a single parameter: the window, the HV settings, and the
momentum.  The average performance using these
non-optimal parameters is labelled ``sub-optimal cluster counting'' in later figures.

\subsection{Cluster Timing for PID}\label{Timing_Results}
\begin{figure}\centering
\includegraphics[width=\columnwidth,clip=true,trim = 0 0 40 30]{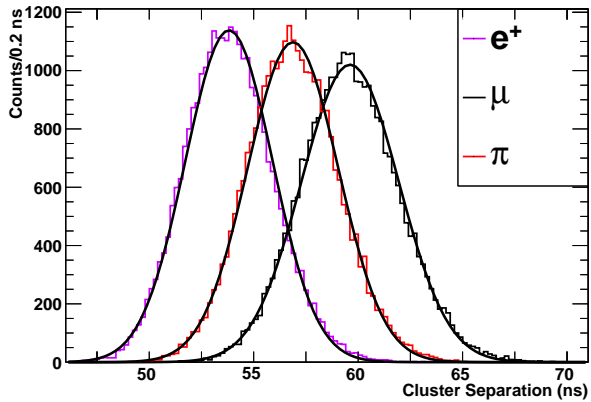}
\caption{Track-wise weighed average of time intervals between
  clusters, in $\unit[50]{ps}$ units for each particle species.  This
  is a run at $\unit[210]{MeV/c}$.  The
  three peaks from left to right are from positrons, pions, and muons,
  respectively}
\label{TrackSepPerSpecies}
\end{figure}

In each cell, we take the average of the time intervals between
consecutive clusters.  In the track composition process, we form a weighted average
of the cell-wise averages, with the weights given by the number of
clusters in each track.  The resulting quantity gives a reasonable
separation for each particle type (Fig.~\ref{TrackSepPerSpecies}). 

Unfortunately the performance is not as good as either the traditional
charge integration or cluster counting (Fig.~\ref{R_and_dEdx}).  In addition, if we form a
tripartite combined likelihood, the improvement relative to the bipartite charge
integration and cluster counting combination is negligible.  Given the
increased computational complexity of calculating the average
separations, it is unlikely that the timing information will be useful
for PID purposes in a real particle physics experiment.

\subsection{Dependence of PID on Gas Gain}\label{GasGainResults}
\begin{figure}\centering
\includegraphics[width=\columnwidth,clip=true,trim = 0 0 40 30]{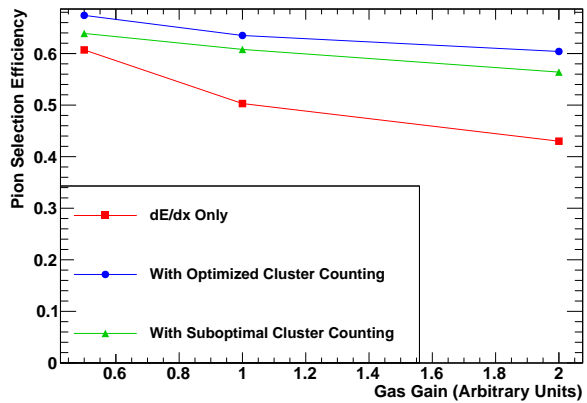}
\caption{Variation in PID performance at the three gas gains that
  were explored.  This is a run at $\unit[210]{MeV/c}$ using an
  inverting $\unit[370]{\Omega}$ amplifier.}
\label{GainVariation}
\end{figure}

The gas gain of the prototypes depends on the choice of sense wire
voltage and on the gas.  We tested only one gas, a mixture of helium
and isobutane in a ratio of $90:10$.  A nominal voltage was selected
as described in Sec.~\ref{Wire_Voltages}.  The actual gas gain for
our gas mix and voltages is on the order of $10^5$, measured offline
using an \ce{^{55}Fe} source.  The procedure aims to
obtain oscilloscope signals with roughly the same amplitude with all
the amplifiers.  The dependence of gas gain on
sense wire voltage is approximately exponential\cite{poenaru1997experimental}.  In our case a $\pm\unit[60]{V}$
change corresponds to a doubling or halving of the gas gain.  The
resulting performance after doubling and halving the gain is shown in Fig.~\ref{GainVariation}.

Previous to the experiment, the intuitive notion was that higher gas
gains would be better, since the signals would stand out more from the
random noise on the chamber wires.  It appears however that this is
not the case and that indeed better PID performance can be obtained at
lower gas gains.  Lower performance at higher gas gains is either due
to gas effects (e.g.\ space charge) or to the amplifiers.  
We did not explore the even lower gains where the
performance is expected to decrease again.  Data runs using other
amplifiers with different gain do show the eventual decrease (Sec.~\ref{CableVariationSection}), so the
optimal voltage is not too far from that shown in
Fig.~\ref{GainVariation} (within $\sim\unit[100]{V}$).

When choosing a gas gain for an experiment the most important features
are more often the tracking performance, ageing issues, and
operational issues.  This is more likely to
influence the choice of specific gain, regardless of the PID
performance.  However, if PID performance is
also highly valued, lower gains should be explored.

\subsection{Momentum}
\begin{figure}\centering
\includegraphics[width=\columnwidth,clip=true,trim = 0 0 40 30]{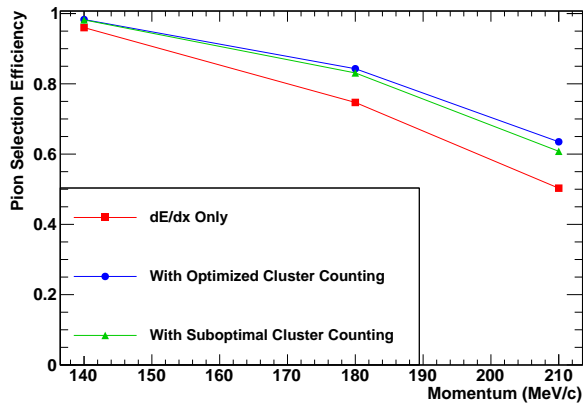}
\caption{Variation in PID performance with momentum.  These three runs
all use the same amplifier with $\unit[370]{\Omega}$ input impedance.}
\label{MomentumVariation}
\end{figure}

As shown in Fig.~\ref{MomentumVariation}, the difference of ionization
between pions and muons is greater at lower momenta.  This is in
agreement with theoretical expectations and simulations.  As expected,
the improvement from adding cluster counting is
most noticeable at the momentum where the overall performance is
worst, making the detector response more uniform.

\subsection{Dependence of PID on Window (Z-position)}\label{Window_Results}
The prototypes have five windows at five thin aluminium positions along their
$\unit[2.7]{m}$ length.  The reference point is chosen to be the
amplifiers, so the high-voltage connectors at the other end of the
chamber are at $\unit[2700]{mm}$.  The centres of the five windows are
$283, 816, 1349, 1883$ and $\unit[2415]{mm}$ from the amplifiers.

Most tests were performed at the windows $1349$ and
$\unit[1883]{mm}$ from the amplifiers, but a sequence of runs was taken
to determine the effect of the signal propagating along the sense
wire.  The sense wire voltages were chosen as described in
Sec.~\ref{Wire_Voltages} at the middle position, but left unaltered for
the other windows in the sequence.  Thus the oscilloscope and
amplifier saturations may change as a function of beam position.

\begin{figure}\centering
\includegraphics[width=\columnwidth,clip=true,trim = 0 0 40 30]{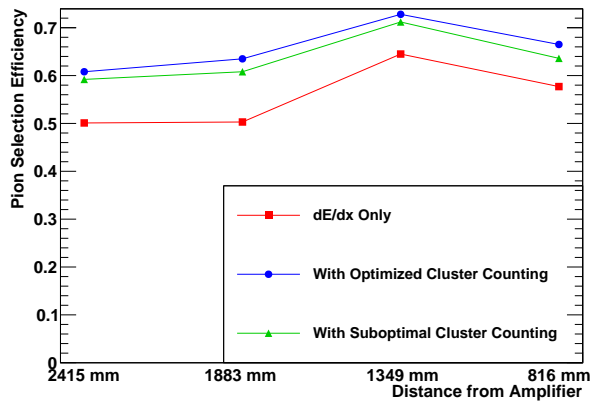}
\caption{Variation in PID performance at the different windows of the
  prototype.  These runs all use the same amplifier with $\unit[370]{\Omega}$ input impedance.}
\label{WindowVariation}
\end{figure}

The tungsten wire is very thin and has a non-negligible DC
resistance ($\unit[421]{\Omega}$ for the $\unit[20]{\mu m}$ diameter wire), so it was expected that the performance would be better at
the windows closer to the amplifiers.  Indeed the runs taken at the
two windows closest to the amplifiers have slightly higher efficiencies
(Fig.~\ref{WindowVariation}) than at the two furthest windows, but the difference is not large.  The variation for this small data set is
also not monotonic, the second-closest window to the amplifiers shows
inexplicably better performance than the closest.  

\subsection{Cables}\label{CableVariationSection}
As mentioned in Sect.~\ref{cables}, we tested two different cable
types, and the effect of adding an additional header
connector to simulate needing to feed through a bulkhead.  Unlike the
previous sections, we did not compare the performance of the
cluster-counting algorithms using parameters optimized on the single
run with non-optimal parameters.  Thus the individual performance
numbers may be optimistic, but the comparison between cable types can
still be done.  
\begin{figure}\centering
\includegraphics[width=\columnwidth,clip=true,trim = 0 20 40 30]{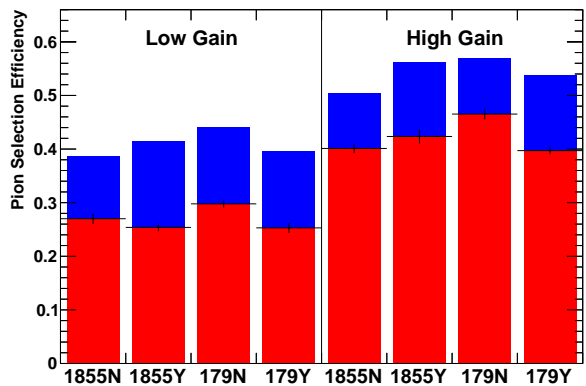}
\caption{Variation in PID performance using different cable types.
  The red is the performance using charge integration only, the blue
  on top
  is the additional performance gain from combining charge integration
  and cluster counting.  All the runs are at $\unit[210]{MeV/c}$ and
  use the same amplifier with $\unit[50]{\Omega}$ input impedance.
  ``Y'' and ``N'' designate the presence or absence of an extra header
  connector.  The first four runs are at low gas gain, and the last
  four are at higher gas gain.}
\label{CableVariation}

\end{figure}
In Fig.~\ref{CableVariation} we show the result from several runs
using an amplifier with $\unit[50]{\Omega}$ input impedance.
The low gain columns have the Chamber $A$ sense wire voltage at
$\unit[1820]{V}$, while the high gain columns are at
$\unit[1835]{V}$.  The high gain voltage was chosen according to
Sec.~\ref{Wire_Voltages}.  Since our gain-doubling voltage is approximately
$\unit[60]{V}$, the low gain columns have about $\unit[84]{\%}$ the
gain of the high gain columns.  The voltages are higher than for the
runs described in the earlier sections because the amplifiers have
different electronic gain.  The cable types 1855 and 179 are
described in Sec.~\ref{cables}, while the Y and N
designations indicate the presence or absence of the extra
header connector, respectively.

A general trend to be noticed is that the high gain columns have
noticeably better performance than the low gain columns, which is
contrary to what was shown in Sec.~\ref{GasGainResults}.  This is
likely because these amplifiers have different electronic gains and
the selected wire voltages do not lie in the same performance region
as the results shown in Sec.~\ref{GasGainResults}.

The cable type and the inclusion of the header connector only
marginally affect the figure of merit, by an amount less than the
typical variation between identical runs and from the track composition
process $\sim\unit[5]{\%}$.  It is tempting to see that the 179N
columns are the highest between the two sets, but the difference is
not nearly as dramatic as the variation due to gas gain or the
additional contribution of cluster counting itself.

\subsection{Amplifiers}\label{Amplifier_Results}
As described in Sec.~\ref{amplifiers}, we tested several types of
amplifiers, mostly distinguished by their input impedance and gain.
We remind the reader that the sense wire voltages used are different
for the various amplifiers, and were chosen to get approximately constant signal
amplitude as described in Sec.~\ref{GasGainResults}.

\begin{figure}\centering
\includegraphics[width=\columnwidth,clip=true,trim = 0 0 40 30]{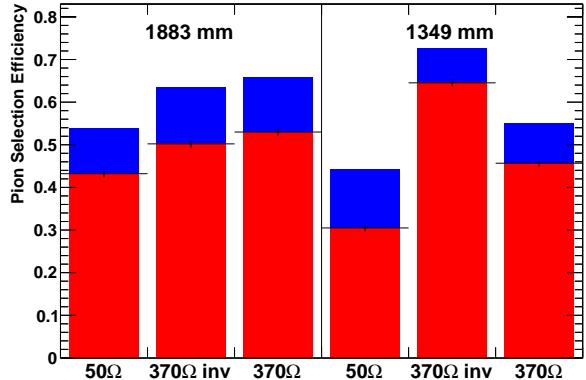}
\caption{Variation in PID performance using different amplifiers.
  The red is the performance using charge integration only, the blue
  is the additional performance gain from combining charge integration
  and cluster counting.  The upper labels indicate the beam position along
  the sense wire, measured from the amplifier.  The ``inv'' label
  indicates an inverting amplifier.}
\label{AmplifierVariation}
\end{figure}

In Fig.~\ref{AmplifierVariation}, the results from three different
amplifiers at two different positions along the sense wire are shown.  The input impedance of
each amplifier is indicated, and the amplifiers with the same labels
are the same for the two different positions.  The $\unit[370]{\Omega}$ ``inv'' amplifier returns
an inverted signal, while the others do not.  

There is considerable variation between the amplifiers, but the
general result is that the $\unit[370]{\Omega}$ amplifiers give the
best results.  This indicates the importance of matching the
amplifier input impedance with the impedance and termination of the
drift chamber itself.  Unfortunately the indication of the best
amplifier is not very strong, as a proper study of the optimal gas
gain for each amplifier was not done in this experiment.  The variation between the amplifiers
in Figure~\ref{AmplifierVariation} is of the same order as the
variation with gas gain for a single amplifier shown in
Figure~\ref{GainVariation}.  It is possible that the variations
seen here are mostly due to gain effects rather than the impedance and
implementation details of the amplifiers.

\subsection{Summary of Results}
The studies undertaken attempt to explore a multidimensional parameter
space, so the results are difficult to summarize concisely.  Here we
restate the lessons learned from each study described above.

The various cluster counting algorithms all perform roughly
equivalently (Sec.~\ref{CC_Results}).  Their parameters must be optimized for good
performance, but the regions of good performance in parameter-space
are quite large.  Even sub-optimal parameters only give slightly worse
performance.  More advanced techniques (such as the timeout booster)
can compensate for a less-optimized algorithm, but are unnecessary
when the algorithm is optimized properly.  

Optimal smoothing for the cluster-counting algorithms is on
the order of a few nanoseconds, indicating that a higher sampling rate
is unnecessary.  The corresponding Nyquist frequency is on the order
of hundreds of $\unit{MHz}$.  This means that the successful implementation of
cluster counting does not depend on getting overly expensive or
customized hardware.  Indeed the best algorithm studied simply applies
a threshold to the second-derivative of the signal, a process that
can be done with analog electronics or in an FPGA. 

Cluster timing gives results that are slightly poorer than cluster
counting used alone (Sec.~\ref{Timing_Results}).  When combined with charge integration and
cluster counting however, the improvement is minor compared to charge
integration and cluster counting without the cluster timing.  Given
the additional complexity of storing and calculating average cluster
timings, this technique is unlikely to be worth exploring further.

PID performance depends strongly on having the proper wire voltages
and thus gas gains (Sec.~\ref{GasGainResults}).  In some configurations, higher gain is not
necessarily better, but this is dependent on the choice of amplifier.
Thus for a given amplifier and equipment configuration,
the optimal gas gain must be carefully determined.  

There is not much variation in PID performance as a function of the
beam position along the sense wire length (Sec.~\ref{Window_Results}).  Since the signal is
attenuated while travelling along the sense wire, this effect is
coupled with the gain of the amplifier and the choice of wire
voltages.  

The choice of cable types and additional connectors seems to have a
negligible effect on the PID performance (Sec.~\ref{CableVariationSection}).  Performance is very
sensitive to the choice of amplifier (Sec.~\ref{Amplifier_Results}), but this is coupled with the
sense wire voltage.  There is a weak indication that matching the
amplifier input impedance with the impedance and termination of the
chamber itself gives better performance. 

\section{Conclusions}
The general result is clear: implementing cluster counting increases
the particle identification capability of a drift chamber.  We make no
claim of having found the optimal equipment and analysis techniques in
the multidimensional parameter space that we explored.  Thus we can
state that cluster counting improves PID performance even in
sub-optimal conditions.  

The absolute improvement in the pion selection efficiency at $\unit[90]{\%}$
muon rejection is generally around $\unit[10]{\%}$ (e.g.\ from
$\unit[50]{\%}$ to $\unit[60]{\%}$, and see Fig.~\ref{R_and_dEdx}).  The improvement is
greatest when the PID performance from charge integration only is
poorest, thus making the detector PID response more uniform.  

The optimal smoothing times for cluster-counting algorithms are on the
order of a few nanoseconds, corresponding to a Nyquist frequency of
hundreds of $\unit{MHz}$.  Thus successful cluster counting can be
accomplished even with modest hardware.

All future particle physics experiments that use a drift chamber for
PID should strongly consider a cluster-counting option.  This study
shows that performance gains can be obtained that justify the
additional complexity and cost of a cluster-counting drift chamber.

\section*{Acknowledgements}
This work was supported by the Natural Sciences and Engineering
Research Council of Canada and TRIUMF.  We thank Jerry Va'vra for
lending us the MCPs for our TOF system, and Hirosiha Tanaka for
lending us the oscilloscope for our Data Acquisition.

We thank Wyatt Gronnemose and Steven Robertson for their assistance during the beam test.

The analysis made extensive
use of open-source software: ROOT\cite{ROOT}, Python\cite{python}, IPython\cite{ipython}, PyROOT\cite{pyroot}
and NumPy\cite{5725236}. 

\begin{appendices}

\section{\mbox{Cluster-Counting Algorithms}}\label{CC_appendix}
Here are contained precise definitions of the cluster-counting and
smoothing algorithms used in this work.  We define a signal or trace
as a series of voltage samples indexed by a discrete time variable
$V(t)$.  Though the time variable has units (in our raw format the
units are $\unit[50]{ps}$), here we treat it as an integer index.  In
general, a signal will have $N$ samples indexed with integer $t$
running from $0$ to $N-1$.

\subsection{Smoothing Procedures}
Two types of smoothing are used in the algorithms.  One involves
replacing each element of the signal by the average of itself and its
neighbours, without reducing the total number of elements.  The other
reduces the total number of elements, and each element's value is the
average of a set of elements in the original signal.
\subsubsection{Boxcar Smoothing}
The so-called ``boxcar smoothing'' with $n$ frames substitutes each
sample with the average of itself and the previous $n-1$ samples.  The
first $n$ samples ($0$ to $n-1$) are a boundary case, replaced simply by $\tilde{V}_n(n)$.
\begin{equation}
\tilde{V}_n(t) = \begin{cases}
\frac{1}{n}\sum\limits_{i=0}^{n-1}V(t-i)& t \ge n,\\
\tilde{V}_n(n)& t < n.
\end{cases}
\end{equation}

\subsubsection{Averaging}\label{averagingsection}
The so-called ``true averaging'' procedure produces a signal with a
reduced number of samples.  For an $n$-frame averaging, the result is a
series of $k = N \div{n}$ voltages (floored division), indexed with
the integer $\bar{t}$ running from $0$ to $k$.  

\begin{equation}\label{averaging}
\bar{V}_n(\bar{t}) = 
\frac{1}{n}\sum\limits_{i=0}^{n-1}V(s+n\bar{t}+i)
\end{equation}
Here, $n$ is the number of samples that are averaged, $s = N\bmod n$,
and $N$ is the total number of samples in the original trace.

This averaging has the potential to ``divide'' cluster signals if the
averaging bin edges lie on top of a cluster (Fig.~\ref{CC_Cut_Quantities}).  Thus it is useful to
also shift the smoothing bins by adding $n\div{2}$ to the argument of
$V$ inside the sum.  If the smoothing is done with and without the
shift, it is less likely that the same cluster will be divided in both
cases, compared to doing the smoothing only one way.

\subsection{Signal above Average}
This algorithm has two parameters: a number of frames for smoothing
and a threshold.  From the non-smoothed signal at time $t$ is subtracted
the $n$-frame smoothed signal at time $t-1$.  If the resulting
quantity crosses the threshold $\Delta$ downwards, a cluster is
identified at that time.

The cluster times found by this algorithm are those $t$ in
$\{\max(n,2)..N\}$ that satisfy
\begin{equation}
\begin{split}
(V(t) - \tilde{V}_n(t-1) &< \Delta) \mbox{ and }\\
(V(t-1) - \tilde{V}_n(t-2) &\ge \Delta).
\end{split}
\end{equation}

\subsection{Smooth and Delay}
This algorithm has four parameters: two smoothing times, a delay, and
a threshold.

Two copies of the original signal are smoothed by different amounts ($p$ and $q$
frames) using the ``boxcar smoothing''.  The $q$-frame smoothed copy
is then delayed by $d$ frames, and the two copies are then
subtracted.  If the resulting quantity crosses the threshold $\Delta$
downwards, a cluster is counted at that time.

The cluster times found by this algorithm are those $t$ in $\{d..N\}$
that satisfy
\begin{equation}
\begin{split}
\frac{\tilde{V}_p(t)-\tilde{V}_q(t-d)}{d} &< \Delta \mbox{ and }\\
\frac{\tilde{V}_p(t-1)-\tilde{V}_q(t-1-d)}{d} &\ge \Delta.
\end{split}
\end{equation}
The ``Signal above Average'' algorithm is a special case with $p = 1$, $q = n$, and
$d = 1$.  Another special case can be constructed with $d = 0$ with the denominator set to $1$.

It can be shown that if the two smoothing times are equal ($p = q$), the quantity computed with smoothing $q$
and delay $d$ is identical to that computed with smoothing $d$ and
delay $q$.  Thus the parameter range can be restricted to $d > q$
without loss of generality.  

\subsection{Second Derivative}
This algorithm has two parameters: a smoothing time and a threshold.
It uses the true averaging procedure rather than the ``boxcar
smoothing'', so the time is labelled $\bar{t}$ as in Sec.~\ref{averagingsection}.  Simply put, the second derivative is calculated and
compared with a threshold.

The second derivative is calculated as follows:
\begin{equation}
\bar{V}''(\bar{t}) = \frac{1}{\delta^2}\Bigl( [\bar{V}(\bar{t}+2)-\bar{V}(\bar{t}+1)] - [\bar{V}(\bar{t}+1) - \bar{V}(\bar{t})]\Bigr)
\end{equation}
where $\delta$ is the time interval corresponding to the $n$ samples
that were averaged to do the smoothing.

The times of the clusters found with the second-derivative algorithm
are those $\bar{t}$ in $\{0..N\div n\}$ that satisfy
\begin{equation}
\bar{V}''(\bar{t}) < \Delta \mbox{ and } \bar{V}''(\bar{t}-1) \ge \Delta.
\end{equation}

Because this algorithm uses the true averaging, it suffers from the
problem of potentially ``dividing'' cluster signals between smoothing
bins (Fig.~\ref{CC_Cut_Quantities}).  Thus we also implemented a two-pass second-derivative algorithm
that looks for clusters a second time on the averaged signal with a
delay applied as described in Sec.~\ref{averagingsection}.  The
numbers of clusters found in each pass are added together.  It is understood that the
resulting cluster count is inflated because many clusters will be
double-counted, but nevertheless it is an appropriate variable for
identifying particles.

\subsection{Timeout Booster}
The so-called ``timeout booster'' takes as an input the list of
clusters found by one of the above algorithms.  It considers these as
cluster candidates, and validates or rejects each one in turn.

For a given cluster candidate, the voltage and time in the original waveform at
which the cluster-finding algorithm was triggered is recorded.  Then
following the waveform forward, the voltage is checked to see when it
has recovered above the recorded value (the pulses are negative). If the voltage
recovered within the timeout window, it is a short-lived pulse and
thus rejected as a fake.  If the timeout is reached without the
voltage recovering, it is long-lived and kept as a real cluster.

For a list of potential clusters ${t_i'}$, real clusters satisfy
\begin{equation}
V(t) < V(t_i') \mbox{ for all } t \mbox{ in } \{t_i'..(t_i'+T)\}
\end{equation}
where $T$ is the chosen timeout.  The rejection of fake clusters by
the timeout procedure permits the use of lower thresholds in the
original algorithm.  The lower threshold increases the efficiency of
finding real clusters (smaller miss rate) but increases the rate of
detecting fake clusters.  The timeout procedure then eliminates most
of the fake clusters, keeping the real ones.

\end{appendices}

\end{document}